\documentclass[pra,showpacs,preprintnumbers,amsmath,amssymb,floatfix]{revtex4}
\usepackage{dcolumn}
\usepackage{bm}

\usepackage{amsmath}    
\usepackage{graphicx}   
\def\be{\begin{equation}}
\def\ee{\end{equation}}
\def\bea{\begin{eqnarray}}
\def\eea{\end{eqnarray}}

\def\oper#1{\hat{\mathbf{#1}}} 

\begin{document}

\title{Inducing complete population oscillations in systems with externally induced dipole moments}
\author{Duje Bonacci}
\affiliation{Physical Chemistry Department, R. Bo\v{s}kovi\'c
Institute, Bijeni\v{c}ka 54, 10000 Zagreb, Croatia}
\email{dbonacci@irb.hr}

\date{\today}

\begin{abstract}
A two level system is considered which has no static dipole
moment, e.g. molecule $H_2$ in its ground electronic state. If
strong enough external field is applied, it will dynamically
distort such a system and supply it with time (and field)
dependent dipole moment. Although it is impossible to do so in the
undistorted system which has no coupling to the dipole component
of the external field, having induced in it a dipole moment, the
rotational and vibrational dynamics of such system can be
manipulated using lasers. In this work, a system is considered in
which the external perturbation dynamically induces the transition
dipole moment between only two distinct levels. The aim of the
work is to show how the driving pulse of the following form:
\begin{equation}
  F[t]= F_{\rm 0} \; m[t] \; \cos{[\omega[t] \; t]}
\label{pulse}
\end{equation}
can be analytically designed, that will produce Rabi-like complete
population oscillations between the two levels.
\end{abstract}

\pacs{3.65.Sq}

\maketitle

\newpage
\section{Calculation}

\subsection{Dynamical equation}

Consider a two level system for which in unperturbed state no
spectroscopically allowed dipole transition exists between its two
levels (e.g. molecule $H_2$). Such system cannot be directly
manipulated using the standard methods of laser control.

However, if such a system is exposed to sufficiently intensive
external perturbation, time-dependent dipole transition moment
between its two states can be dynamically induced. This dipole
transition moment can then be used to manipulate the system's
internal (rotational, vibrational, ...) dynamics.

It can be demonstrated \cite{balint-kurti2004} that the dynamics
of such two-level system subjected to strong coherent external
radiation is governed by the following equation:
\begin{eqnarray}
\label{2 level}
  \frac{d}{d t}
    \begin{pmatrix}
      c_\alpha [t] \\
      c_\beta [t] \\
    \end{pmatrix}
    = \frac{d F[t]}{d t}
    \frac{\mu_{\alpha \beta}[t]}{\omega_{\alpha \beta}[t]}
    \begin{pmatrix}
     - i\; \frac{\mu_{\alpha \alpha}[t]}{\mu_{\alpha \beta}[t]}\; \omega_{\alpha \beta}[t]\; t
     & e^{i\; s_{\alpha \beta}[t]\;\omega_{\alpha \beta}[t]\; t}
     \\ -e^{-i\; s_{\alpha \beta}[t]\; \omega_{\alpha \beta}[t]\; t}
     & - i\; \frac{\mu_{\beta \beta}[t]}{\mu_{\alpha \beta}[t]}\; \omega_{\alpha \beta}[t]\; t
     \\
    \end{pmatrix}
    \begin{pmatrix}
      c_\alpha [t] \\
      c_\beta [t]  \\
    \end{pmatrix}
\end{eqnarray}

In this equation: $c_{\alpha,\beta}[t]$ are the wave-function
expansion coefficients; $F[t]$ is the external field, $\mu_{ij}[t]
\equiv \langle i | \hat{\mu} | j \rangle $ is the field-induced
dipole moment of state or dipole transition moment ($\hat{\mu}$ is
the dipole moment operator); $\omega_{\alpha \beta} \equiv
|E_\alpha -E_\beta |$ is the absolute value of the instantaneous
transition frequency between the two states of the system; and
$s_{\alpha \beta} \equiv Sign[E_\alpha -E_\beta]$ is the sign of
that transition frequency.

\subsection{The external field}
The external field is constructed so as to fit the following
analytic form:
\begin{equation}
  F[t]= F_{\rm 0}\; m[t]\; \cos{[\omega[t] \; t]}
\label{pulse}
\end{equation}

Here, $F_{\rm 0}$ is the maximum amplitude of the field achieved
throughout the pulse, $m[t]$ is the pulse envelope $(0\le m[t]\le
1)$ and $\omega[t]$ is the time-dependent frequency of the
external perturbation. All of these three parameters are freely
tunable.

It should be noted that such 'analytic' form needs not be fully
analytic - pulse envelope and pulse chirp can be purely numerical
functions. Also, such seemingly rigid form of the pulse puts no
serious restriction on the possible variety of actual pulse shape.
Indeed, by varying the three pulse parameters, any reasonable
pulse shape can be constructed. The only important caveat is that
pulse must be such that at any time the field varies much more
rapidly than the envelope. If this is granted, it
straightforwardly follows that:

\begin{eqnarray}
\label{pulse derivation} \frac{d F[t]}{d t} &\approxeq& - F_{\rm
0}\;m[t]\; \omega[t] \; \sin{[\omega[t]\; t]} \nonumber \\
&=& i\; F_{\rm 0}\; m[t]\; \omega[t]\; \frac{e^{i\; \omega[t]\;
t}-e^{-i\; \omega[t]\; t}}{2}
\end{eqnarray}

\subsection{Transforming the time variable}

Next, for further analytic calculation convenience, the following
transformations and shorthands are introduced:

\begin{eqnarray}
d\tau &\equiv& \frac{F_{\rm 0}\; m[t]\; \omega[t]}{2}
\frac{\mu_{\alpha \beta}[t]}{\omega_{\alpha \beta}[t]}\; d t \\
\Delta_{\pm}[\tau] &\equiv& \omega[t[\tau]] \pm \omega_{\alpha
\beta}[t[\tau]]\\
f_i [\tau] &\equiv& 2\; \frac{\mu_{i i}[t]}{\mu_{\alpha
\beta}[t]}\; \omega_{\alpha \beta}[t]\; t\; sin[\omega[t[\tau]]\
t[\tau]]\ ; \ (i= \alpha, \beta)\\
\begin{pmatrix}
  a_\alpha [t] \\
  a_\beta [t]  \\
\end{pmatrix} &\equiv&
\begin{pmatrix}
  c_\alpha [t[\tau]] \\
  c_\beta [t[\tau]]  \\
\end{pmatrix}
\end{eqnarray}

With Eq. (\ref{pulse derivation}) included, in terms of these
quantities the dynamical equation Eq. (\ref{2 level}) becomes:

\begin{eqnarray}
\label{non RWA}
  \frac{d}{d t}
    \begin{pmatrix}
      a_\alpha [\tau] \\
      a_\beta [\tau] \\
    \end{pmatrix}
    &=&\\
     -&i& \begin{pmatrix}
     - f_{\alpha}[\tau]
     & s_{\alpha \beta}[t[\tau]](e^{-i\; s_{\alpha \beta}[t]\; \Delta_{-}[t]\;
     t}- e^{i\; s_{\alpha \beta}[t]\; \Delta_{+}[t]\; t})
     \\ s_{\alpha \beta}[t[\tau]](e^{i\; s_{\alpha \beta}[t]\; \Delta_{-}[t]\;
     t}- e^{-i\; s_{\alpha \beta}[t]\; \Delta_{+}[t]\; t}) &
     - f_{\beta}[\tau]
      \nonumber \\
    \end{pmatrix}
    \begin{pmatrix}
      a_\alpha [\tau] \\
      a_\beta [\tau]  \\
    \end{pmatrix}
\end{eqnarray}

\subsection{Introduction of RWA and Rabi oscillations condition}
Under certain conditions (see e.g. \cite{bonacci2003.1}) whose
validity can be checked retrospectively, rotating wave
approximation can be introduced to simplify the obtained full
dynamical equation of the system, Eq. (\ref{non RWA}). If RWA is
valid, then the dynamical significance of the rapidly rotating
elements in the dynamical matrix (the ones containing
$\Delta_+[\tau]$) is negligible, and these elements can be dropped
from all further calculations. Hence, approximate dynamical
equation is obtained:

\begin{eqnarray}
\label{RWA}
  \frac{d}{d t}
    \begin{pmatrix}
      a_\alpha [\tau] \\
      a_\beta [\tau] \\
    \end{pmatrix}
    = -i \begin{pmatrix}
     - f_{\alpha}[\tau]
     & s_{\alpha \beta}[t[\tau]]\; e^{-i\; s_{\alpha \beta}[t]\; \Delta_{-}[t]\;
     t}
     \\ s_{\alpha \beta}[t[\tau]]\; e^{i\; s_{\alpha \beta}[t]\; \Delta_{-}[t]\;
     t} &
     - f_{\beta}[\tau]\\
    \end{pmatrix}
    \begin{pmatrix}
      a_\alpha [\tau] \\
      a_\beta [\tau]  \\
    \end{pmatrix}
\end{eqnarray}
Now the final unitary transformation is sought:
\begin{equation}
\label{transformation}
  \begin{pmatrix}
  b_\alpha(\tau) \\
  b_\beta(\tau) \\
  \end{pmatrix}
  = e^{-i \oper{\Lambda} (\tau)}
  \begin{pmatrix}
  a_\alpha(\tau) \\
  a_\beta(\tau) \\
  \end{pmatrix}
\end{equation}
with:
\begin{eqnarray}
  \oper{\Lambda} [\tau] =
  \begin{pmatrix}
  \rho_1 [\tau] & 0 \\
  0 & \rho _2[\tau] \\
  \end{pmatrix}.
\end{eqnarray}
where $\rho_1[\tau]$ and $\rho_2[\tau]$ are freely adjustable
functions, such that the final transformed system vector
satisfies:

\begin{equation}
\label{standard rabi}
  \frac {d }{d \tau}
  \begin{pmatrix}
  b_\alpha[\tau] \cr
  b_\beta[\tau]
  \end{pmatrix}
  = \pm i
  \begin{pmatrix}
  0 & 1 \cr
  1 & 0 \cr
  \end{pmatrix}
  \begin{pmatrix}
  b_\alpha[\tau] \cr
  b_\beta[\tau]
  \end{pmatrix}
\end{equation}

As Eq.(\ref{standard rabi}) is identical to the standard form of
Rabi oscillations equation in a two-level system (for details, see
e.g. \cite{demtroeder1988}), the corresponding solution would
represent complete population transfer oscillations between the
two levels. Introducing transformation (\ref{transformation}) into
(\ref{RWA}), the following equation is obtained:

\begin{eqnarray}
\label{RWA and Rabi}
  \frac{d}{d t}
    \begin{pmatrix}
      b_\alpha [\tau] \\
      b_\beta [\tau] \\
    \end{pmatrix}
    = -i \begin{pmatrix}
     - f_{\alpha}[\tau] + \frac{d}{d\tau}\rho _1[\tau]
     & s_{\alpha \beta}[t[\tau]]\; e^{-i (s_{\alpha \beta}[t[\tau]]\; \Delta_{-}[\tau]
     + (\rho _1 [\tau]- \rho _2[\tau]))\; t[\tau]}
     \\ s_{\alpha \beta}[t[\tau]]\; e^{i (s_{\alpha \beta}[t[\tau]]\; \Delta_{-}[\tau]
     + (\rho _1 [\tau]- \rho _2[\tau]))\; t[\tau]} &
     - f_{\beta}[\tau] + \frac{d}{d\tau}\rho _2[\tau]\\
    \end{pmatrix}
    \begin{pmatrix}
      b_\alpha [\tau] \\
      b_\beta [\tau]  \\
    \end{pmatrix}
\end{eqnarray}

If this is to be fitted to form (\ref{standard rabi}), the
following conditions must be fulfilled:
\begin{eqnarray}
  \frac{d}{d\tau} {\rho _1}(\tau)&=& f_{\alpha}[\tau] , \\
  \frac{d}{d\tau} {\rho _s}(\tau)&=& f_{\beta}[\tau], \\
  s_{\alpha \beta}[\tau]\; \Delta_{-}[\tau]\; t[\tau] + (\rho _1 [\tau]- \rho_2[\tau])&=& 0 ,
\end{eqnarray}
which can be compactly written as:
\begin{equation}
\label{difjed}
  \frac{d}{d \tau} (\Delta_{-}[\tau]\; t[\tau])=-s_{\alpha \beta}[\tau](f_{\alpha}[\tau]-f_{\beta}[\tau])
\end{equation}

\subsection{Analyticaly optimized frequency chirp}
Integrating Eq. (\ref{difjed}) and reverting to the original time
coordinate $t$ yields the recurrent formal solution for the
optimized driving frequency $\omega[t]$:
\begin{equation}
\omega[t]=\omega_{\alpha \beta}[t]-s_{\alpha \beta}[t]
\frac{F_{\rm 0}}{t} \int_{t_0}^{t} (\mu_{\alpha
\alpha}[t_1]-\mu_{\beta \beta}[t_1])\; m[t_1]\; \omega[t_1]\;
t_1\; sin[\omega[t_1]\; t_1]\; dt_1
\end{equation}
As this is recurrent equation, id does not provide directly the
optimized frequency. However, with numerical computational power
nowadays available it should be a fairly simple and quick task to
obtain the solution using some a rather simple computer iteration
scheme.

Hence, using this solution, fully controlled and complete
population oscillations can be induced in the strongly perturbed
system.



\section*{Acknowledgment}
I am very grateful to Prof. Gabriel Balint-Kurti for insightful
discussion and sharing of some ideas from his own work. These
provided both indispensable sparks and firm guidelines in the
development of the results presented herein.


\end{document}